\documentclass[a4,11pt]{article}
\title{{\em Ab initio} statistical mechanics of surface adsorption \\
and desorption: II. Nuclear quantum effects}
\author{D. Alf\`{e}$^{1,2,3,4}$ and M. J. Gillan$^{2,3,4}$ \\
$^1$Dept. of Earth Sciences, UCL, Gower St, London WC1E 6BT, UK \\
$^2$London Centre for Nanotechnology, UCL, Gordon St, London WC1H 0AH, UK \\
$^3$Dept. of Physics, UCL, Gower St, London WC1E 6BT, UK \\
$^4$Thomas Young Centre at UCL, Gordon St, London WC1H 0AH, UK
}
\usepackage{a4wide}
\usepackage{graphicx}

\begin{document}
\maketitle
\abstract{
We show how the path-integral formulation of quantum statistical
mechanics can be used to construct practical {\em ab initio} techniques 
for computing the chemical potential of molecules adsorbed on surfaces, with
full inclusion of quantum nuclear effects. The techniques we describe
are based on the computation of the potential of mean force on a chosen
molecule, and generalise the techniques developed recently for
classical nuclei. We present practical calculations based on
density functional theory with a generalised-gradient
exchange-correlation functional for the case
of H$_2$O on the MgO~(001) surface at low coverage. We note that
the very high vibrational frequencies of the H$_2$O molecule
would normally require very large numbers of time slices (beads)
in path-integral calculations, but we show that this requirement
can be dramatically reduced by employing the idea of thermodynamic
integration with respect to the number of beads. The validity and
correctness of our path-integral calculations on the H$_2$O/MgO~(001)
system are demonstrated by supporting calculations on a set of
simple model systems for which quantum contributions to the free
energy are known exactly from analytic arguments.
}


\section{Introduction}
\label{sec:intro}

In a recent paper~\cite{alfe2007}, referred to here as paper~I,
we described an {\em ab initio} scheme for calculating
the chemical potential and other thermodynamic properties of
systems of adsorbed molecules, with full inclusion of entropy effects.
We presented illustrative calculations based on density functional
theory (DFT) for H$_2$O at low coverage on the MgO~(001) surface,
and showed how the chemical potential, and hence the desorption rate
can be calculated without statistical-mechanical approximations, apart
from the treatment of the nuclei as classical particles.
That initial study was intended as a first step in developing
{\em ab initio} methods for calculating the thermodynamic properties
of adsorbate systems with better than chemical accuracy, i.e. better
than 1~kcal/mol. As a contribution to these developments, we present
here a practical scheme for including quantum nuclear effects in the
{\em ab initio} statistical mechanics of adsorbates, and we report
here some results for H$_2$O on MgO~(001) at low coverage.

Quantum effects are likely to have a very significant effect on the
properties of some adsorbates, particularly those containing protons,
such as H$_2$O. The zero-point energies $\frac{1}{2} \hbar \omega$ of
the symmetric and antisymmetric stretching modes and the bending mode
of the H$_2$O molecules are 227, 233 and 99~meV respectively (5.23,
5.37 and 2.28~kcal/mol)~\cite{tennyson2001}. Since partial or complete
dissociation of adsorbed H$_2$O occurs quite 
commonly~\cite{langel1994,giordano1998,lindan1998,schaub2001,yu2003}, 
it is clear
that inclusion of quantum nuclear effects will sometimes be essential
for accurate calculations. Needless to say, the future achievement of
chemical accuracy will also depend crucially on the development of
{\em ab initio} methods which go beyond conventional DFT, and that is
currently the object of intensive 
research~\cite{li2008,ma2009,muller2009,paulus2009}. 
However, the availability
of improved {\em ab initio} methods is not our immediate concern here.

We emphasised in paper~I that we aim to avoid statistical-mechanical
approximations, and in particular we 
do not allow ourselves to resort to harmonic approximations,
since these will not be satisfactory for the disordered adsorbates
that are of ultimate interest. 
The overall strategy we shall present here closely resembles that of
paper~I, the main new feature being that the
path-integral formulation of quantum mechanics is used to treat the
nuclei. We noted in paper~I that the chemical
potential determines the ratio between the surface adsorbate density
and the gas-phase density with which it is in equilibrium, and that
this ratio can be expressed as an integral over the normalised density
distribution $y ( z )$ of molecules as a function of height $z$ above
the surface. Since $y ( z )$ can be expressed in terms of a potential
of mean force $\phi ( z )$, it can be computed from a sequence of
thermal-equilibrium simulations in each of which the mean force on a
chosen molecules is calculated, with the height $z$ of this molecule
above the surface being constrained to have a fixed value. As we shall
see, these relationships remain true in path-integral theory.

Path-integral statistical mechanics is based on the well known
isomorphism between a quantum system of $N$ particles and a classical
system of $N$ cyclic chains, each consisting of $P$ beads (time
slices), with neighbouring beads in each chain being coupled by harmonic
springs~\cite{feynman1965,feynman1972,parrinello1984,berne1986,gillan1990,
marx1996,tuckerman1996}. The meaning of this is 
that each cyclic chain represents the
propagation of an individual quantum particle in imaginary time $\tau$
from $\tau = 0$ to $\tau = \beta \hbar$. The ratio
between the densities of adsorbed and gas-phase molecules, and the
expression for this ratio in terms of a probability distribution $y (
z )$ retain their validity when the nuclei are described by quantum
mechanics, and we shall show that $y ( z )$ can be expressed in terms
of a potential of mean force $\phi ( z )$ which can be derived from
appropriately constrained simulations on the system of cyclic chains
given by the path-integral isomorphism. In this sense, the
generalisation of the methods of paper~I to include
quantum nuclear effects is in principle straightforward.

We shall use the example of H$_2$O on MgO~(001) to show that it is 
practically feasible to calculate the chemical potential of an adsorbate
using {\em ab initio} path-integral methods. The illustrative
calculations are for the case of low coverage, i.e. isolated molecules,
because we are not yet able to perform the statistical sampling at
higher coverages. For isolated H$_2$O molecules on MgO~(001), the
changes of vibrational frequencies on adsorption are not large, so
that quantum nuclear effects do not lead to great changes of the
chemical potential. Nevertheless, the technical challenges of the
calculations are interesting and substantial, because the high
vibrational frequencies mean that the number of path-integral beads 
needs to be very high to achieve the required accuracy. To meet this
challenge, we had to develop a new technique, in which thermodynamic
integration is performed with respect to the number of beads. A significant
part of the paper is concerned with the tests we have developed
to demonstrate that our path-integral techniques work correctly.

The rest of the paper is organised as follows. The techniques themselves
are laid out in Sec.~\ref{sec:techniques}: 
we start with a brief reminder of the
methods developed in paper ~I, and a summary of basic path-integral theory;
then we explain how this theory can be used to create a path-integral
generalisation of paper~I that correctly yields the chemical
potential with full inclusion of quantum nuclear effects; techniques
for improving the computational efficiency and for performing 
thermodynamic integration with respect to the number of beads are
also outlined; at the end of Sec.~\ref{sec:techniques}, 
we note the DFT techniques
used in our practical calculations. Sec.~\ref{sec:testing} describes the tests
we have performed on the path-integral machinery; these tests consist
of simulations on a set of harmonic vibrational models involving
the H$_2$O molecule acted on by various external fields, for which
the results can be checked against values known from analytical
formulas. The calculations on the chemical potential of the H$_2$O
molecule adsorbed at finite temperature on MgO~(001) are
presented in Sec.~\ref{sec:full_calcs}. Finally, the implications of the work
and the prospects for future applications and developments
are discussed in Sec.~\ref{sec:discussion}.


\section{Techniques}
\label{sec:techniques}


\subsection{Chemical potential of adsorbed molecules: classical approximation}
\label{sec:classical_theory}

The classical theory of paper~I is expressed in terms
of the spatial distribution $\bar{\rho} ( z )$ of molecules when there
is full thermal equilibrium between the gas phase and the
adsorbate. This distribution is defined so that $\bar{\rho} ( z ) \, d
{\bf r}$ is the probability of finding a molecule in volume element $d
{\bf r}$ at height $z$ above the surface, averaged over positions in
the plane parallel to the surface. The detailed form of $\bar{\rho} (
z )$ depends on which point within the molecule (e.g. centre of mass,
position of a specified atom in the molecule, etc.)  is chosen 
to construct the distribution; we refer to this as the 
``monitor point''~\cite{alfe2007}.  
In practice, it is convenient to work with the
normalised function $ y ( z ) \equiv \bar{\rho} ( z ) / \rho_0$, where
$\rho_0$ is the mean density of molecules in the gas phase. The ratio
of the density $\sigma$ of adsorbate molecules (number per unit area
on the surface) and the gas-phase density $\rho_0$ is:
\begin{equation}
\sigma / \rho_0 = \int_{- \infty}^{z_0} dz \, y ( z ) \; ,
\label{eqn:int_y}
\end{equation}
where $z_0$ is the height above the surface below which molecules
are counted as being ``adsorbed''. This ratio is also related to the
excess chemical potentials of gas-phase and adsorbed molecules:
$\sigma / \rho_0 = d \exp [ \beta \Delta \mu^\dagger ( \sigma , T ) ]$, where
$\Delta \mu^\dagger \equiv \mu_{\rm gas}^\dagger - \mu_{\rm ads}^\dagger$
is the difference of excess chemical potentials, defined in paper~I,
$\beta = 1 / k_{\rm B} T$ and $d$ is a length chosen for the
purpose of defining $\mu_{\rm ads}^\dagger$. We refer to 
$\Delta \mu^\dagger$ as the ``excess chemical potential difference'' (ECPD).
In principle, the choice of $z_0$ affects $\sigma$, and hence
$\Delta \mu^\dagger$; but in most situations of interest $y ( z )$
for $z < z_0$ is enormously greater than the gas-phase value
$y ( z \rightarrow \infty ) = 1$, so that in practice the dependence
of $\sigma / \rho_0$ on $z_0$ can be neglected. For the same reason,
the dependence of $\sigma / \rho_0$ and $\Delta \mu^\dagger$ on the choice
of monitor point is extremely weak.

Since $y ( z ) >> 1$ in the adsorption region $z < z_0$, the computation
of $y ( z )$ is a ``rare event'' problem.
To overcome this problem, $y ( z )$ is expressed as $y ( z
) = \exp ( - \beta \phi ( z ) )$, where the potential of mean force
$\phi ( z )$ represents the free energy of the entire system when the
height of the molecule above the surface is constrained to have value
$z$, with the condition $\phi ( z \rightarrow \infty ) = 0$.  To
compute $\phi ( z )$, we use the fact that $\langle F_z \rangle_z = -
d \phi / d z$ is the $z$-component of the thermal average force acting
on the molecule when its height is constrained to be $z$. In the
practical calculations of paper~I, {\em ab initio} molecular dynamics (m.d.)
simulation was used to perform the constrained statistical sampling in
the canonical ensemble, and hence the computation of $\langle F_z
\rangle_z$. Integration of $\langle F_z \rangle_z$ then yields
$\phi ( z )$ and hence $y ( z )$. Related techniques
have been developed by other groups (see e.g. Ref.~\cite{fichthorn2002}).


\subsection{Basic path-integral methods}
\label{sec:pi_basic}

We summarise briefly the path-integral background needed in this work;
for details, the many reviews should be
consulted~\cite{feynman1965,feynman1972,berne1986,gillan1990,marx1996,
tuckerman1996}. The Helmholtz free
energy $F$ of a general system at temperature $T$ is $F = - k_{\rm B}
T \ln Z$, where the partition function is $Z = {\rm Tr} \, [ \exp ( -
\beta \hat{H} ) ]$, with $\beta = 1 / k_{\rm B} T$ and $\hat{H} =
\hat{T} + \hat{V}$ the Hamiltonian of the system. Here $\hat{T}$ is
the operator for the sum of kinetic energies of all the nuclei, and
$\hat{V} = U ( \{ {\bf r}_i \} )$ in the present work represents the
electronic ground-state energy of the system when the nuclei are fixed
at positions ${\bf r}_i$ ($i = 1, 2, \ldots N$).

As usual in path-integral theory, $Z$ is rewritten as
$Z = {\rm Tr} \, \left[ ( \hat{\rho} ( \beta / P ) )^P \right]$,
where $\hat{\rho} ( \beta / P ) = \exp ( - ( \beta / P ) \hat{H} )$
is the propagator for imaginary time $\beta \hbar / P$, 
which is approximated by the Trotter short-time formula:
\begin{equation}
\hat{\rho} ( \beta / P ) \simeq  e^{- \beta \hat{V} / 2 P}
\hat{\rho}_0 ( \beta / P ) e^{- \beta \hat{V} / 2 P} \; ,
\end{equation}
with $\hat{\rho}_0 ( \beta / P )$ the free-particle propagator,
given in coordinate representation by:
\begin{equation}
\rho_0 ( {\bf R} , {\bf R}^\prime ; \beta / P ) \equiv
\langle {\bf R} | \hat{\rho}_0 ( \beta / P ) | {\bf R}^\prime \rangle = 
A_P \exp \left[
- \sum_{i=1}^N \frac{m_i P}{2 \beta \hbar^2} 
| {\bf r}_i - {\bf r}_i^\prime |^2 \right] 
\; .
\end{equation}
Here, ${\bf R}$ and ${\bf R}^\prime$ are points in the configuration
space of the whole system, specified by the collections of
positions $\{ {\bf r}_i \}$ and $\{ {\bf r}_i^\prime \}$, and the
prefactor $A_P$ is given by:
\begin{equation}
A_P = \prod_{i=1}^N \left( \frac{m_i P}{2 \pi \beta \hbar^2} \right)^{3/2} \; ,
\end{equation}
with $m_i$ the mass of nucleus $i$.
The resulting approximation to the partition function, denoted
here by $Z_P$, is then:
\begin{equation}
Z_P = \left( A_P \right)^P
\int \prod_{s=0}^{P-1} \prod_{i=1}^N d {\bf r}_{i s} \,
\exp [ - \beta ( T_P ( \{ {\bf r}_{i s} \} ) +
V_P ( \{ {\bf r}_{i s} \} ) ] \; ,
\label{eqn:Z_P}
\end{equation}
where ${\bf r}_{i s}$ is the position of nucleus $i$ at time-slice $s$,
and $T_P$ and $V_P$ are given by:
\begin{eqnarray}
T_P ( \{ {\bf r}_{i s} \} ) & = & \sum_{s=0}^{P-1} \sum_{i=1}^N
\frac{1}{2} \kappa_{P , i} 
| {\bf r}_{i \, s+1} - {\bf r}_{i s} |^2 \nonumber \\
V_P ( \{ {\bf r}_{i s} \} ) & = & 
\frac{1}{P} \sum_{s=0}^{P-1} U ( \{ {\bf r}_{i s} \} ) \; .
\label{eqn:T_and_V}
\end{eqnarray}
The spring constants $\kappa_{P , i}$ are given by
$\kappa_{P , i} = m_i P / \beta^2 \hbar^2$.
The formula for $Z_P$ expresses the widely used approximation
to the partition function of the $N$-particle quantum system in
terms of the partition function of a classical
system of $N$ cyclic chains, each consisting of $P$ ``beads'' (time slices),
with neighbouring beads
in each cyclic chain being coupled by harmonic springs. The exact
partition function is recovered in the limit $P \rightarrow \infty$.

The spatial distributions of the positions of beads belonging to
a particular time-slice in the classical system of cyclic chains
have a special significance, because they are equal (in the
$P \rightarrow \infty$ limit) to the corresponding equal-time
spatial distributions in the quantum system. For example, the
probability density of finding a particular nucleus at a particular
spatial point ${\bf r}$ in the quantum system is equal to the
probability density of finding a chosen bead on the cyclic chain
representing that nucleus at point ${\bf r}$. This means that
appropriate potentials of mean force in the classical cyclic-chain
system allow us to calculate spatial probability distributions in
the quantum system.


\subsection{Constraints and mean force in path-integral scheme}
\label{sec:constraints}

In the case treated later of H$_2$O adsorbed on MgO~(001), we could
choose to take as monitor point the position of the O atom in the
H$_2$O molecule. In that case, the normalised height distribution $y (
z )$ of this monitor point in the quantum system is equal to the
height distribution $y ( z )$ of a chosen bead on the cyclic chain
representing that O atom in the classical cyclic-chain system.
Alternatively, if we took as monitor point the centre of mass of the
H$_2$O molecule, i.e. the position ${\bf r}^{\rm cm} = ( m_{\rm O}
{\bf r}_1 + m_{\rm H} {\bf r}_2 + m_{\rm H} {\bf r}_3 ) / ( m_{\rm O}
+ 2 m_{\rm H} )$ in an obvious notation, then the height distribution
$y ( z )$ of the $z$-component $z^{\rm cm}$ of this position in the
quantum system is equal to the height distribution $y ( z )$ of the
quantity $z^{\rm cm}$ constructed from the O and H positions, all at
the same chosen time-slice in the classical cyclic-chain system. Just
as in the classical theory of paper~I, the detailed
form of $y ( z )$ will depend on the choice of monitor point, but the
integral $\int_{- \infty}^{z_0} dz \, y ( z )$ will not depend 
appreciably on this choice (see above). In the quantum case,
there are also other possible ways of constructing monitor points. One
possibility is that we monitor the centre of mass of the centroids of
the cyclic paths of the O and the two H atoms of the H$_2$O
molecule. (The centroid of the cyclic chain of nucleus $i$ is the
average ${\bar{\bf r}}_i = P^{-1} \sum_{s=1}^P {\bf r}_{i s}$.)  If we
do this, then $y ( z )$ for the classical system of cyclic chains does
not necessarily correspond to any obervable of the quantum system.
Nevertheless, the integral $\int_{- \infty}^{z_0} dz \, y ( z )$ will
still be the same as for other choices of monitor point. This choice
of ``centroid centre of mass'' is the one that we use in the present
work. 

For any of the foregoing choices of monitor point, the distribution
$y ( z )$ can be computed from the potential of mean force
$\phi ( z )$, defined in the usual way as
$y ( z ) = \exp ( - \beta \phi ( z ) )$, and the derivative
$- d \phi / d z$ is $\langle F_z \rangle_z$, the mean value of the
$z$-component of the force on the monitor point, when the $z$-component
of the position of this monitor point is constrained to have a particular
value.


\subsection{Thermal sampling}
\label{sec:thermal_sampling}

When performing constrained thermal-equilibrium simulations of the
cyclic chain system in order to calculate quantities such as $\langle
F_z \rangle_z$, we can choose to use any simulation method that
samples configuration space in the required manner. As in
paper~I, we use molecular dynamics (m.d.) simulation with
Nos\'{e}~\cite{nose1984} and Andersen~\cite{andersen1980} 
thermostats to perform sampling
according to the canonical ensemble. We noted in paper~I that with
this approach we are free to choose the m.d. masses of the particles in any
way that helps to improve the sampling efficiency, since spatial
averages do not depend on these masses. When we use m.d. to perform
thermal sampling of the system of cyclic chains, it is therefore
essential to distinguish between the real nuclear masses $m_i$
appearing in the spring constants $\kappa_i$ from the m.d.
``sampling masses'' employed to sample the configuration space.  In
paper~I, the sampling masses were chosen as $M_{\rm H} = 8$, $M_{\rm 
  O} = 16$ and $M_{\rm Mg} = 24$~a.u.

It is well known that if m.d. simulation is used to perform thermal
sampling in the path-integral technique, with each bead on every chain
being give a chosen mass, then the sampling efficiency is very poor,
because of the very wide frequency spread of the vibration modes of
each chain. The method we use to overcome this problem, which resembles
the techniques used by other workers (see e.g. Ref.~\cite{ivanov2003}), 
employs a mass matrix on each
cyclic chain, so that the Hamiltonian governing the m.d. evolution is:
\begin{eqnarray}
H =  
\frac{1}{2} \sum_{i=1}^N \sum_{s=0}^{P-1} \sum_{t=0}^{P-1}
{\bf p}_{i s} \cdot {\bf A}_{i \, s t} \cdot {\bf p}_{i t} 
& + & \frac{1}{2} \sum_{i=1}^N \sum_{s=0}^{P-1}
\kappa_i | {\bf r}_{i \, s+1} - {\bf r}_{i \, s} |^2 \nonumber \\
& + & P^{-1} \sum_{s=0}^{P-1} U ( {\bf r}_{1 s} \, \ldots {\bf r}_{N s} ) \; ,
\end{eqnarray}
where ${\bf A}_{i \, s t}$ are the elements of the (positive definite)
inverse mass matrix for the cyclic chain representing nucleus $i$.
Each mass matrix is chosen so that the vibration frequencies of all
modes of each chain are similar, and so that the total mass of each chain
has a value that would be suitable for the sampling mass in
the classical system. This is achieved if we take:
\begin{equation}
A_{i \, s t} = \frac{1}{M_i} \sum_{n=0}^{P-1}
\frac{(C_i / P ) \cos [ 2 \pi n ( s - t ) / P ]}
{(C_i / P ) + 2 \kappa_i ( 1 - \cos ( 2 \pi n / P ) )} \; .
\end{equation}
The total sampling mass $M_i$ of each chain is chosen exactly as
in the classical simulations of paper~I (see above). The spring 
constants $C_i$ are chosen so that $( C_i / M_i )^{1/2}$
is a typical frequency characterising the dynamics of nucleus $i$ in
the classical simulations performed with masses $M_i$.


\subsection{Switching number of beads}
\label{sec:switch_beads}

In some of our calculations, it is helpful to obtain the free
energy by successive approximations. First, we calculate
$F_P = - k_{\rm B} T \ln Z_P$ with a number of
beads $P$ which is known to be insufficient. We then obtain
a more accurate value $F_{2 P}$ by adding the difference
$F_{2 P} - F_P$. If the accuracy is still insufficient,
we repeat the process. There is a simple
way of using thermodynamic integration to calculate the difference
$F_{2 P} - F_P$, which we now explain.

We define a generalised form of the free energy $F_{2 P}$ for the system
having $2 P$ time slices by introducing weights $w_s$
($s = 0, \ldots 2 P - 1$) into the definition of the potential
energy $V$ of eqn~(\ref{eqn:T_and_V}):
\begin{equation}
V_{2 P} ( \{ {\bf r}_{i s} \} ) = \frac{1}{2 P} \sum_{s=0}^{2P-1}
w_s U ( \{ {\bf r}_{i s} \} ) \; .
\end{equation}
The partition function $Z_{2 P}$ and the free energy $F_{2 P}$ are
now functions of the $w_s$. We choose the $w_s$ to have the form 
$w_s = 1 + \lambda$ for even $s$ and $w_s = 1 - \lambda$ for odd $s$,
so that the partition function $Z_{2 P}$ is given by:
\begin{equation}
Z_{2 P} = 
\left( A_{2 P} \right)^{2 P} 
\int \prod_{s=0}^{2P-1} \prod_{i=1}^N d {\bf r}_{i s} \,
\exp \left[ - \beta \left( 
T_{2 P}  ( \{ {\bf r}_{i s} \} ) +
\frac{1}{2 P} \sum_{s=0}^{2P-1} ( 1 + ( -1 )^s \lambda ) 
U ( \{ {\bf r}_{i s} \} )  
\right) \right] \; .
\label{eqn:F2P}
\end{equation}
Clearly, $Z_{2 P}$ and hence the free energy $F_{2 P}$ 
depend on $\lambda$.
From eqn~(\ref{eqn:F2P}), we readily obtain a formula for the variation of
$F_{2 P}$ with $\lambda$:
\begin{equation}
\partial F_{2 P} / \partial \lambda =
 - \left\langle 
\frac{1}{2 P} \sum_{s=0}^{P-1}
\left(
U ( \{ {\bf r}_{i \, 2s+1} \} ) - U ( \{ {\bf r}_{i \, 2s} \}
\right)
\right\rangle_\lambda \; .
\label{eqn:deriv_F2P}
\end{equation}

Now the quantity $F_{2 P} ( \lambda = 0 )$ is just the usual
approximation to the free energy calculated with $2 P$ beads, as is
clear from eqn~(\ref{eqn:F2P}). 
On the other hand, $F_{2 P} ( \lambda = 1 )$ is identical
to $F_P ( \lambda = 0 )$, which is the normal free energy calculated
with $P$ beads. To see this, we note from eqn~(\ref{eqn:F2P}) that
for $\lambda = 1$ the potential term in the Boltzmann factor becomes:
\begin{equation}
\frac{1}{2 P} \sum_{s=0}^{2P-1} ( 1 + ( - 1 )^s )
U ( \{ {\bf r}_{i s} \} ) =
\frac{1}{P} \sum_{s=0}^{P-1} U ( \{ {\bf r}_{i \, 2s} \} ) \; ,
\label{eqn:lambda1potl}
\end{equation}
which is the potential term that appears in $F_P ( \lambda = 0 )$. But
this means that we can integrate over all the positions ${\bf r}_{i s}$ for
odd $s$, since these no longer appear in the potential term. In doing
the integrations, we use the fact that:
\begin{eqnarray}
\left( A_{2 P} \right)^2 \int \prod_{i=1}^N d {\bf r}_{i \, 2s+1} 
\! \! \! \! \! & \, & \! \! \! \! \!
\exp \left[
- \beta \sum_{i=1}^N \frac{1}{2} \kappa_{2P, i}
| {\bf r}_{i \, 2s+1} - {\bf r}_{i \, 2s} |^2 \right]
\exp \left[
- \beta \sum_{i=1}^N \frac{1}{2} \kappa_{2P , i}
| {\bf r}_{i \, 2s+2} - {\bf r}_{i \, 2s+1} |^2 \right] = \nonumber \\
&  \, & A_P \exp \left[
- \beta \sum_{i=1}^N \frac{1}{2} \kappa_{P , i}
| {\bf r}_{i \, 2s+2} - {\bf r}_{i \, 2s} |^2 \right] \; ,
\label{eqn:convol}
\end{eqnarray}
which expresses the fact that 
$\hat{\rho}_0 ( \beta / ( 2 P ) ) \hat{\rho}_0 ( \beta / ( 2 P ) ) =
\hat{\rho}_0 ( \beta / P )$.
On using eqns~(\ref{eqn:lambda1potl}) and (\ref{eqn:convol}) in 
eqn~(\ref{eqn:F2P}), we obtain
$F_{2 P} ( \lambda = 1 ) = F_P ( \lambda = 0 )$.

It follows from this, and from eqn~(\ref{eqn:deriv_F2P}), that:
\begin{equation}\label{eqn:F2P-FP}
F_{2 P} - F_P = \int_0^1 d \lambda \,
\left\langle \frac{1}{2 P} \sum_{s=0}^{P-1}
( U ( \{ {\bf r}_{i \, 2s+1} \} ) - U ( \{ {\bf r}_{i \, 2s} \} ) 
\right\rangle_\lambda \; .
\end{equation}
This is the formula we use to calculate accurate values of free energy
by successive approximation.


\subsection{{\em Ab initio} techniques}
\label{sec:ab_initio}

Our practical simulations, like those of paper~I, were made
with the projector-augmented-wave implementation of 
DFT~\cite{blochl1994,kresse1999}, using
the VASP code~\cite{kresse1996}. The DFT computations on all
the $P$ path-integral images of the system (i.e. beads, or time-slices)
are performed in parallel, with the operations for each image also
distributed over groups of processors (typically, there are 16 processors
allocated to each image for our full calculations on the MgO slab
with or without H$_2$O). As in paper~I, we use the PBE exchange-correlation
functional~\cite{perdew1996}, with a plane-wave cut-off of 400~eV and
an augmentation-charge cut-off of 605~eV. 


\section{Testing the path-integral machinery}
\label{sec:testing}

To verify that the path-integral techniques work correctly, we have
used them to calculate the free energy changes when various external
potentials are applied to an isolated H$_2$O molecule in free space.
We have designed these potentials so that they mimic the changes that
occur when a molecule passes from free space to the adsorbed state:
the formation of vibrational and librational modes from free
translations and rotations, and the alteration of intramolecular
vibration modes. The potentials are designed so that the changes
of free energy are known exactly. These tests also give us information
about the number of beads $P$ needed to obtain results of specified
accuracy. All the tests are performed with appropriately modified
versions of the VASP code, with the technical settings mentioned
above (Sec.~\ref{sec:ab_initio}).


\subsection{Vibration of centre of mass}
\label{sec:c_of_m}

To mimic the conversion of free translations to vibrations, we
introduce a potential that acts only the centre of mass of the molecule.
This potential leaves molecular rotations and internal vibrations
completely unchanged.
We use the notation ${\bf r}_i$ for the positions of the three atoms
in the molecule, with $i = 1$, 2, 3 corresponding to the O atom and
the two H atoms respectively. The position of the physical centre of
mass is ${\bf r}^{\rm cm} \equiv ( m_{\rm O} {\bf r}_1 +
m_{\rm H} ( {\bf r}_2 + {\bf r}_3 ) ) / M$, where $M = m_{\rm O} + 2 m_{\rm H}$
is the total mass of the molecule. The total energy
$U_0 ( {\bf r}_1 , {\bf r}_2 , {\bf r}_3 )$ of the molecule in free space
depends only on relative positions, such as ${\bf r}_2 - {\bf r}_1$ and
${\bf r}_3 - {\bf r}_1$, since it is translationally invariant.
We add to $U_0$ an external
potential $U_{\rm ext} ( z^{\rm cm} )$ depending only on
the $z$-coordinate of the centre of mass. We want $U_{\rm ext} ( z^{\rm cm} )$
to consist of a harmonic well of spring constant $\alpha$ and depth
$V_0$, and we want it to go to zero as $z \rightarrow \infty$. To obtain
this, we adopt the form:
\begin{eqnarray}
U_{\rm ext} ( z ) & = & - V_0 + \frac{1}{2} \alpha ( z - z_0 )^2 \; ,
\; \; \; \; \; z < z_1 \nonumber \\
& = & - \frac{1}{2} \beta ( z - z_2 )^2 \; , \; \; \; \; \; 
z_1 < z < z_2 \nonumber \\
& = & 0 \; , \; \; \; \; \; z_2 < z \; .
\end{eqnarray}
The constants are chosen so that $U_{\rm ext}$ and its first
derivative are continuous at $z = z_1$. In this model, the rotational
and internal-vibrational properties of the molecule are completely
unchanged, and the only effect of $U_{\rm ext}$ is to produce harmonic
oscillations of the centre of mass in the $z$-direction when $z <
z_1$. We take the potential parameters and $T$ to be such that, when
the molecule is in thermal equilibrium in the potential well, the
probability of finding the centre of mass outside the region $z < z_1$
is negligible. Then the values of $\Delta \mu^\dagger$ in the
quantum and classical cases are exactly 
$k_{\rm B} T \ln [ 2 \sinh ( \hbar \omega / 2 k_{\rm B} T ) ]$
and $k_{\rm B} T \ln ( \hbar \omega / k_{\rm B} T )$, so that their
difference (quantum minus classical) is $k_{\rm B} T \ln [ \sinh (
\hbar \omega / 2 k_{\rm B} T ) / ( \hbar \omega / 2 k_{\rm B} T ) ]$,
where $\omega = \sqrt{\alpha / M }$ is the oscillation frequency 
of the centre of mass.

In the classical case, simulations are superfluous, since if the 
$z$-component of the centre of mass position is constrained to be
$z$, then the force on the centre of mass is simply
$F_z = - d U_{\rm ext} / d z$. In the quantum case, the coordinate
we choose to constrain is the $z$-component of the centre of mass
of the centroids ${\bar{z}}^{\rm cm} \equiv m_{\rm O} {\bar{z}}_1 +
m_{\rm H} ( {\bar{z}}_2 + {\bar{z}}_3 )) / M$.
Now it is readily shown that for a perfectly harmonic system in
thermal equilibrium, the probability distribution of the path centroid
is identical to that of the corresponding classical system. This implies
that the mean force is also identical, and that the potentials of
mean force in the quantum and classical cases can differ by
at most a constant. In the present case of a piecewise harmonic
$U_{\rm ext}$, the mean force is therefore the same in the quantum
and classical systems, except for those ${\bar{z}}^{\rm cm}$ values
for which beads are distributed on both sides of the ``join'' position
$z_1$. Contributions to the quantum-classical difference of
$\Delta \mu^\dagger$ therefore come only from this join region.

We have performed practical tests on this model for a variety
of parameter values. An example is the case $\alpha = 31.975$~eV/\AA$^2$,
which gives a centre-of-mass oscillation frequency of 20.8~THz. With
$| z_1 - z_0 | = 0.5$~\AA\ (we take $\beta = \alpha / 25$), the centre
of mass is almost completely confined to the region
$z < z_1$ in both the quantum and classical cases at $T = 100$~K.
In this example, the exact quantum-classical difference of
$\Delta \mu^\dagger$ is 23.2~meV. Our path-integral calculations
of mean force for this case, performed with $P = 8$ (spot checks
with $P = 16$ showed no appreciable differences), gave a difference
of $23.2 \pm 0.5$~meV, in perfect agreement with the exact value.


\subsection{Libration of normal to molecular plane}
\label{sec:normal_to_plane}

To illustrate the conversion of free rotations to librations, we
take a potential that acts on the vector normal to the molecular
plane of the H$_2$O molecule. A suitable vector is:
\begin{equation}
{\bf Q}_\perp = ( {\bf r}_2 - {\bf r}_1 ) \times ( {\bf r}_3 - {\bf r}_1 ) \; ,
\end{equation}
and we define the potential as:
\begin{equation}
V_{\rm ext} = \frac{1}{2} \alpha ( n_x^2 + n_y^2 ) \; ,
\end{equation}
where the unit vector ${\bf n}$ is ${\bf Q}_\perp / | {\bf Q}_\perp |$.
For $\alpha = 0$, the molecule rotates freely, but for large positive $\alpha$
the normal executes small oscillations about the $z$-axis. If $\alpha$
is large enough, these oscillations are harmonic, and their
frequencies are $\omega_1 = ( \alpha / I_1 )^{1/2}$ and
$\omega_2 = ( \alpha / I_2 )^{1/2}$, where $I_1$ and $I_2$ are
the moments of inertia for rotations about the two symmetry axes
in the molecular plane. In-plane rotations (moment of inertia $I_3$)
are unaffected. The free energy in the presence of $V_{\rm ext}$
is then the free energy of the two librations plus that of the
free in-plane rotation. For the free energy in the absence of $V_{\rm ext}$,
we use the standard expression for the free energy of an 
asymmetric top~\cite{stripp1951}.
The free energy increase $\Delta F$ caused by switching on $V_{\rm ext}$
is thus known almost exactly. For our tests, we take $\alpha = 2.00$~eV
and $T = 100$~K; under these conditions, the two librational modes
are almost exactly in the quantum ground state. The free energy
increase is $\Delta F = 111$~meV.

We test the path-integral methods using a series of simulations with
scaled external potential $\lambda V_{\rm ext}$, with $\lambda$ going
from 0 to 1. The thermal average of $V_{\rm ext}$ in the path-integral
system was calculated using $\lambda$ values of 0.0, 0.03, 0.06,
0.125, 0.25, 0.5 1.0, the reason for taking closely spaced $\lambda$
values near $\lambda = 0$ being that the thermal average varies
rapidly with $\lambda$ in this region (see
Table~\ref{tab:vext}). Numerical integration gives $\Delta F = 102 \pm
1$~meV, and $\Delta F = 106 \pm 1$~meV using $P = 8$ and $P = 16$
respectively. The small free energy difference obtained between $P =
8$ and $P = 16$ was also reproduced by directly calculating it using
eqn~(\ref{eqn:F2P-FP}), which gives $\Delta F_{16} - \Delta F_8 = 5 \pm
1$~meV. Using eqn~(\ref{eqn:F2P-FP}), we also obtained $\Delta F_{32} -
\Delta F_{16} = 1.5 \pm 0.5$~meV, so that our best estimate for
$\Delta F$ is $ 108.5 \pm 2$~meV, which agrees with the exact value
within the statistical errors.

\begin{table}
\begin{tabular}{c|cc}
\hline
  & \multicolumn{2}{c}{$P$} \\
$\lambda$ & 8 & 16 \\
\hline
0.0   & $680\pm 11$ & $664\pm 12$ \\  
0.03  & $405\pm 12$ & $443\pm 15$ \\  
0.06  & $256\pm 8$ & $247\pm 10$ \\  
0.125 & $153\pm 4$ & $144\pm 5$ \\  
0.25  & $97\pm 2$   & $101\pm 3$ \\
0.5   & $65\pm 1$   & $70\pm 2$ \\    
1.0   & $41\pm 1$   & $47\pm 1$ \\
\hline
$\Delta F$ & $102\pm 1$ & $106 \pm 1$ \\   
\hline
\end{tabular}
\caption{Thermal average $\langle V_{\rm ext} \rangle_\lambda$ as
function of $\lambda$ and free energy change $\Delta F$. Units are
meV.}\label{tab:vext}
\end{table}

\subsection{Alteration of asymmetric stretch frequency}
\label{sec:asym_stretch}

Surface adsorption will alter the intramolecular vibration frequencies,
and to mimic this effect we apply an external potential having the form:
\begin{equation}
Z_{\rm ext} = \frac{1}{2} \kappa \xi^2 \; ,
\end{equation}
where $\xi = ( {\bf r}_2 + {\bf r}_3 - 2 {\bf r}_1 ) \cdot 
( {\bf r}_3 - {\bf r}_2 )$. Clearly, this has no effect on translations
or rotations. By symmetry, $\xi$ remains zero in the presence of symmetric
stretch and bond-angle oscillations. The only mode that is affected
by $Z_{\rm ext}$ is thus the asymmetric stretch mode. In the absence of
$Z_{\rm ext}$, our DFT calculations give the frequency
$\omega_{\rm a} / 2 \pi = 114.8$~THz. We have done tests with the
value $\kappa = 5.0236$~eV/\AA$^4$, which gives the increased
frequency $\omega_{\rm a} / 2 \pi = 151.2$~THz. At $T = 100$~K, this
vibrational mode is in its quantum ground state, so that the
free energy change caused by $Z_{\rm ext}$ is
$\Delta F = \frac{1}{2} \hbar 
( \omega_{\rm a} - \omega_{\rm a}^0 ) = 77.3$~meV.

In our path-integral simulations, we calculate the thermal average of
$Z_{\rm ext}$ as a function of $\lambda$ in the presence of the
scaled potential $\lambda Z_{\rm ext}$, with $\lambda$ going from 0 to 1.
These calculations were done with several different numbers of time-slices
$P$, going from 8 to 96. The computed results for the free energy
change $\Delta F$ (Table~\ref{tab:zext}) show that $P = 96$ gives essentially
exact results, and $P = 64$ is also acceptable, but smaller values give
seriously inaccurate results. This indicates that if we wished to
calculate the chemical potential of adsorbed H$_2$O by brute force 
path-integral simulation, we would have to use at least $P = 64$,
which would be exceedingly expensive.

\begin{table}
\begin{tabular}{cc}
\hline
$P$ & $\Delta F$(meV) \\
\hline
96 & $78 \pm 1.5$ \\
64 & $75 \pm 1.5$ \\
32 & $59 \pm 1.5$ \\
16 & $41 \pm 1$  \\
8  & $21 \pm 1.5$\\
\hline
\end{tabular}
\caption{Free energy of the water molecule (in meV) in the external potential
$Z_{\rm ext}$ at 100 K calculated using path integral simulations with
different values of $P$.}\label{tab:zext}
\end{table}

This gives us an opportunity to test our technique of thermodynamic
integration with respect to $P$. To apply this, we first calculate the
free energy change as $\lambda$ goes from 0 to 1, using $P = 8$.  The
resulting value of $\Delta F$ is in error by nearly 50~meV
(Table~\ref{tab:zext}).  To correct this, we now calculate the free
energy changes $F ( P ) - F ( P / 2 )$, both for the free molecule and
for the molecule acted on by $Z_{\rm ext}$.
The resulting correction gives us $\Delta F ( 8 ) +
\Delta F ( 8 \rightarrow 64 ) = 70 \pm 6$~meV, which agrees within
statistical error with the directly calculated value $\Delta F ( 64 ) =
75 \pm 1.5$~meV.

\subsection{Other tests}
\label{sec:other_tests}

As a further test of librational effects, we have compared path-integral
calculations with quasi-exact results for the case where potentials
act both on the normal to the molecular plane and on the direction
of the molecular bisector. This is achieved by adding to the potential
$V_{\rm ext}$ of Sec.~\ref{sec:normal_to_plane} a 
potential $W_{\rm ext}$ defined by:
\begin{equation}
W_{\rm ext} = \frac{1}{2} \gamma q_x^2 \; ,
\end{equation}
where $q_x$ is the $x$-component of the unit vector given by
${\bf q} = ( {\bf r}_2 + {\bf r}_3 - 2 {\bf r}_1 ) /
| {\bf r}_2 + {\bf r}_3 - 2 {\bf r}_1 |$. The tests were done with
$\gamma = 2.00$~eV, which gives a frequency of $\sim 16$~THz for
in-plane librations. The potential $W_{\rm ext}$ also induces
a small change in the frequency of the asymmetric stretch mode.
As before, the free energy change caused by switching on 
$V_{\rm ext} + W_{\rm ext}$ agrees within statistical error
with the quasi-exact value.

In addition to all these tests on the {\em ab initio} free molecule,
we have also made a completely different kind of test involving
PMF calculations on the molecule as it is brought onto the 
surface of the rigid MgO slab (see Appendix). The methods are the same
as those used for the full path-integral calculations presented
in the next Section.


\section{Full path-integral calculations for H$_2$O on MgO~(001)}
\label{sec:full_calcs}

The PMF techniques of Sec.~\ref{sec:techniques} are now 
applied to the full calculation
of the ECPD $\Delta \mu^\dagger \equiv \mu_{\rm gas}^\dagger -
\mu_{\rm ads}^\dagger$, both with and without nuclear quantum
effects. We denote the difference of $\Delta \mu^\dagger$ values
(quantum minus classical) by $( \Delta \mu^\dagger )_{\rm qc}$. We
know from the tests of Sec.~\ref{sec:testing} that 
in order to obtain values for $(
\Delta \mu^\dagger )_{\rm qc}$ that are correct to better than $\sim
5$~meV, the number of beads $P$ must be no less than 64. Our strategy
is to perform all PMF calculations with $P = 8$, and then to correct
the results by performing thermodynamic integration with respect to
the number of beads, for the free molecule, for the slab-molecule
system, and for the clean slab. All the parameters of the simulated
system were set exactly as in paper~I: we use a
3-layer $2 \times 2$ slab (16 ions per layer, for a total of 48 ions per
repeating unit of the MgO slab), with a vacuum gap of 12.7~\AA, a
lattice parameter of 4.23~\AA\ (the PBE bulk equilibrium value at $T =
0$~K), and $\Gamma$-point electronic sampling.  We use a time-step of
1~fs, and m.d. sampling masses of 24.3, 16.0 and 8.0~a.u.  for Mg, O and H, in
conjunction with the mass-matrix technique 
(Sec.~\ref{sec:thermal_sampling}). With these
parameters, the conservation of the constant of motion was of the same
quality as that obtained in the classical simulation of paper~I. The
calculations were performed at $T = 100$~K, with an Andersen
thermostat~\cite{andersen1980}, and 
constrained simulations were made at 13 different
values of height $z$. We find that simulations of 20~ps suffice to
give a statistical error on the PMF at its minimum of less than 6~meV,
which, as shown in paper~I (Appendix A), is approximately equal to the
error in the ECPD $\Delta \mu^\dagger$.

\begin{figure}
\centerline{
\includegraphics[width=4.0in]{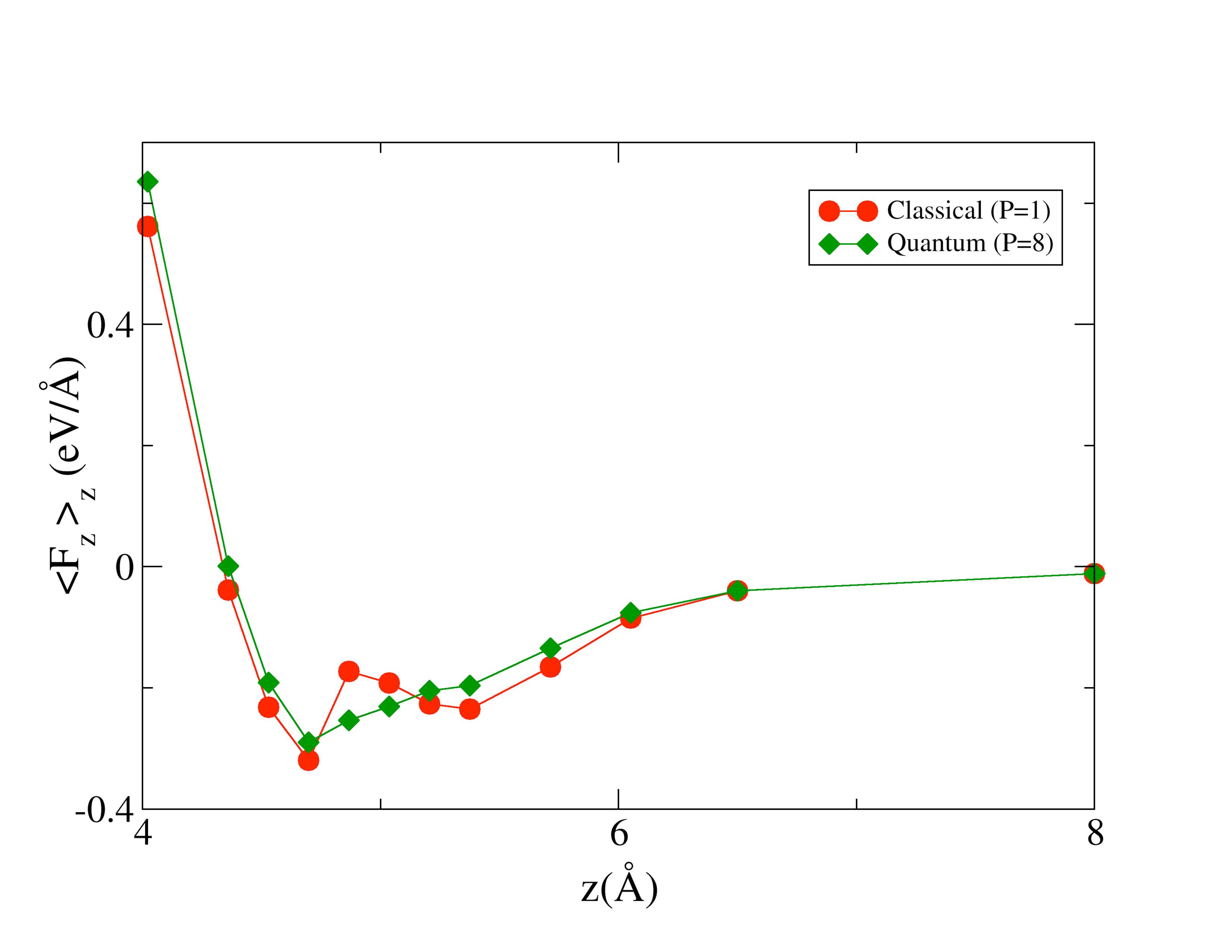}
\includegraphics[width=4.0in]{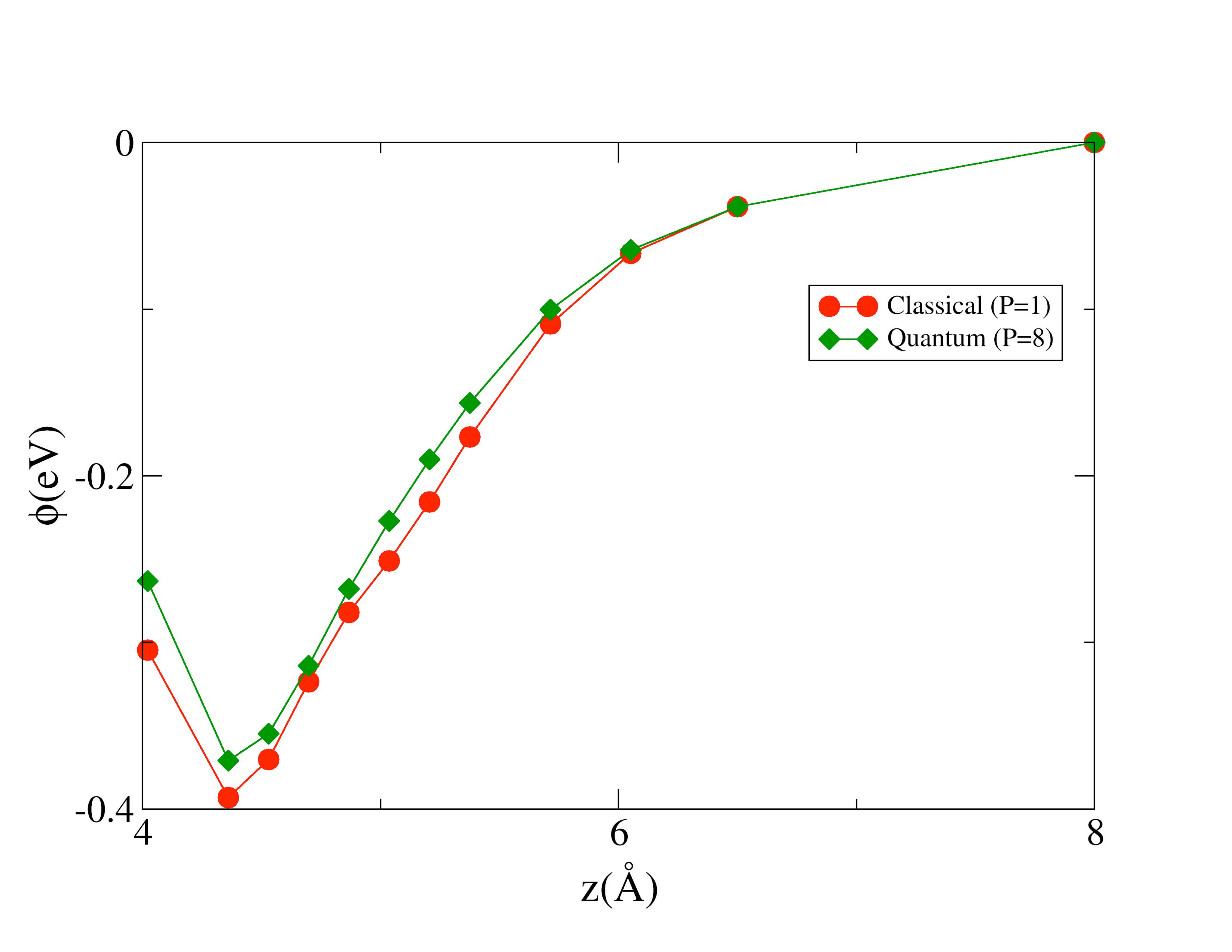}
}
\caption{ Mean force $\langle F_z \rangle_z$ and potential of mean
  force $\phi ( z )$ of H$_2$O on MgO~(001) calculated here using
  path-integral simulation with $P = 8$ beads (diamonds), compared
  with the classical results of paper~I (circles). Units are \AA\ and
  eV. Statistical error bars are smaller than the size of the points.
}
\label{fig:mean_force}
\end{figure}

The mean forces $\langle F_z \rangle_z$ and the PMFs $\phi ( z )$
calculated with $P = 8$ and $P = 1$ (classical case) are shown in
Fig.~\ref{fig:mean_force}. We see that quantum effects are small, but
are easily detectable within the statistical errors. In particular, we
notice that the classical mean force is slightly lower than the
quantum value almost everywhere, apart from the region $4.8 \le z \le
5.2$~\AA, where it is higher. By integrating the PMFs, we obtain a
quantum-classical difference of EPCD of $( \Delta \mu^\dagger )_{\rm
  qc} = - 22 \pm 7$~meV.

\begin{figure}
\centerline{
\includegraphics[width=2.6in]{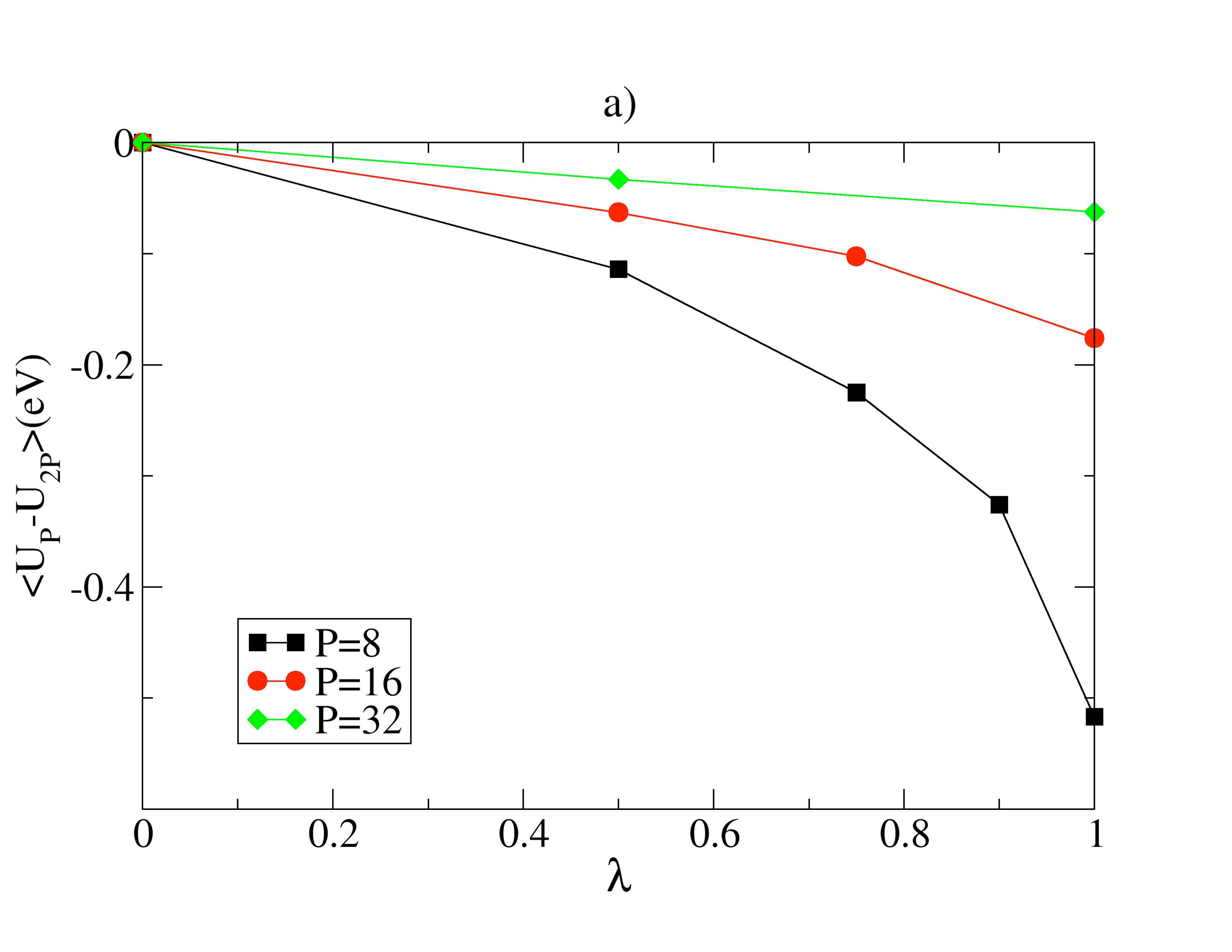}
\includegraphics[width=2.6in]{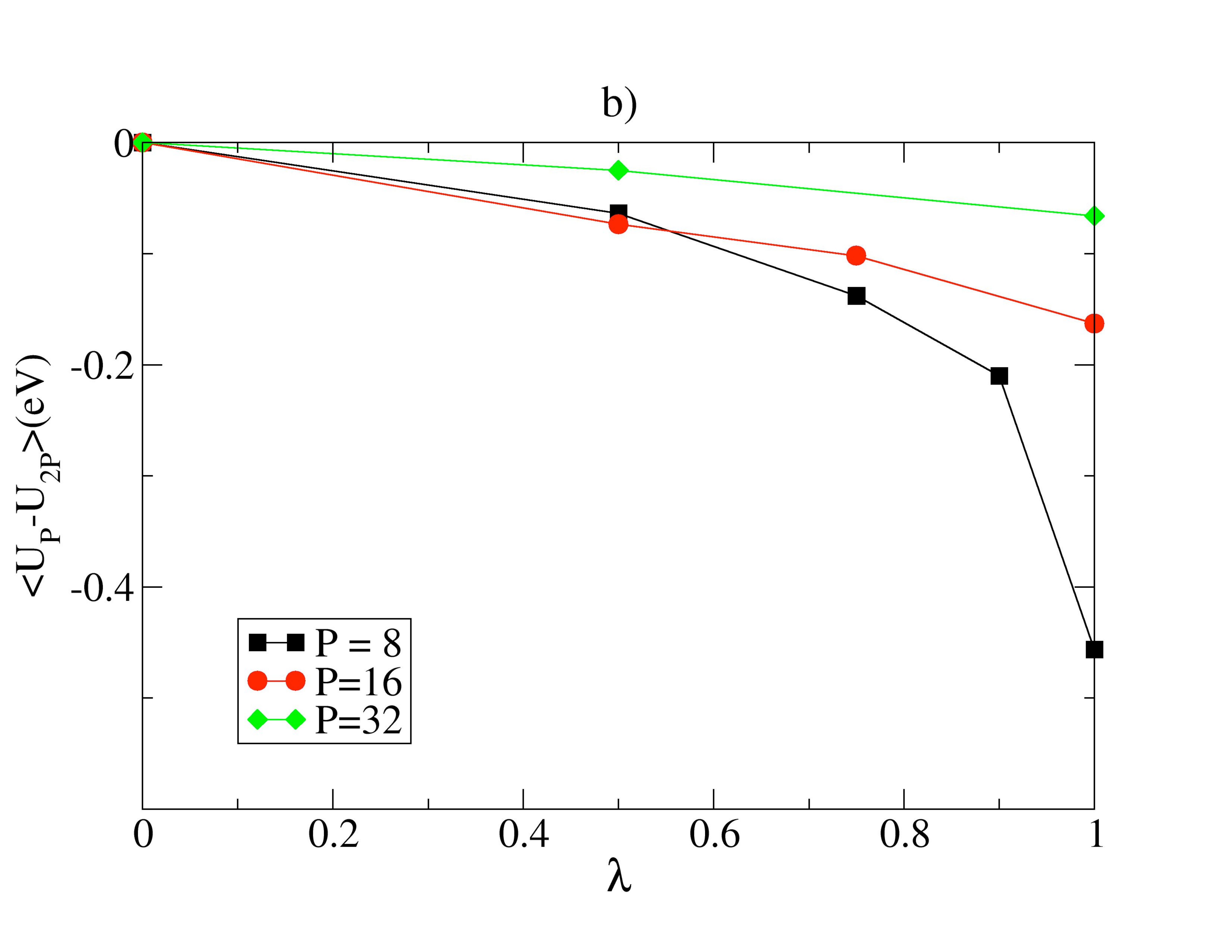}
\includegraphics[width=2.6in]{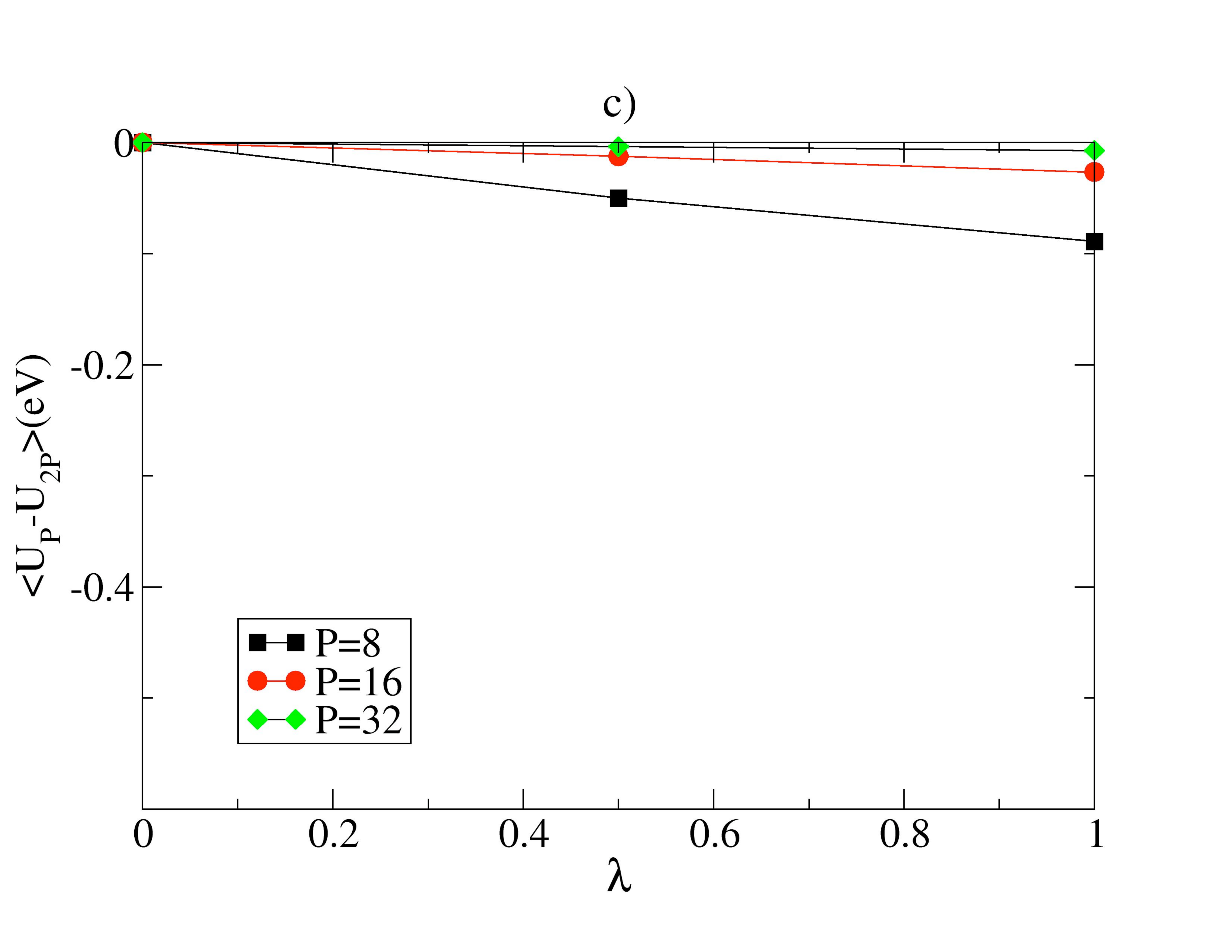}
}
\caption{ Computed results for the quantity $\langle U_P - U_{2 P}
  \rangle_\lambda$ (eV units) used to calculate the difference $F_P -
  F_{2 P}$ between path-integral free energies for numbers of beads
  $P$ and $2 P$. Panels (a), (b) and (c) show results for the
  slab-molecule system, the free molecule, and the bare slab
  respectively. Results for $P = 8$, 16 and 32 are shown by squares,
  circles and diamonds. Statistical error bars are smaller than the
  size of the points.  }
\label{fig:u_of_lambda}
\end{figure}

In computing the free energy changes on going from $P = 8$ to $P = 64$
for the isolated molecule and slab-molecule system, we used 4, 3 and 2
values of $\lambda$ for the $8 \rightarrow 16$, $16 \rightarrow 32$
and $32 \rightarrow 64$ contributions (simulations at $\lambda = 0$
are unnecessary, since $\langle U_P - U_{2P} \rangle_{\lambda = 0} =
0$).  For the clean slab, we needed only two values of $\lambda$,
since the $P = 8 \rightarrow 64$ corrections for this system are
rather small. In Fig.~\ref{fig:u_of_lambda}, we show the values of
$\langle U_P - U_{2P} \rangle_\lambda$ as a function of $\lambda$ for
the three systems. We see that most of the nuclear quantum
contribution comes from the step $8 \rightarrow 16$, but the other two
contributions are not negligible. However, when we combine the
contributions from the three systems, the three contributions $8
\rightarrow 16$, $16 \rightarrow 32$ and $32 \rightarrow 64$ are all
very small, and the total change of $\Delta \mu^\dagger$ (64 beads
minus 8 beads) amounts to only $- 8 \pm 3$~meV.

Including now the correction due to going from $P = 8$ to 64 beads,
the final result for the nuclear quantum correction to the ECPD
comes to $( \Delta \mu^\dagger )_{\rm qc} = - 30 \pm 8$~meV.


\section{Discussion and conclusions}
\label{sec:discussion}

The main purpose of this work has been to establish the practical
possibility of calculating the chemical potential of adsorbed
molecules using {\em ab initio} methods with full inclusion of
nuclear quantum effects, but without resorting to harmonic
approximations. We have shown that within the path-integral
formulation of quantum mechanics this can be achieved by a
generalisation of the {\em ab initio} methods that have already
been used for the case of classical nuclei. In particular, the method
based on computing the mean force acting on a single molecule in a 
series of constrained simulations generalises naturally to the
path-integral framework. Our practical calculations on the chemical
potential of H$_2$O on MgO~(001) at low coverage show that available
computer power suffices to reduce statistical errors to
$\sim 10$~meV, if required. However, this is achievable only
thanks to a techniques we have introduced for performing thermodynamic
integration with respect to the number of beads (time-slices). 

Naturally, future calculations of this kind will be of greatest interest
when the nuclear quantum effects are large, and this will typically
be true when very high intra-molecular vibrational frequencies
are strongly affected by surface adsorption. Many examples of this
are known, among which are water adsorption on both oxide and
metal surfaces. For example, it has recently been shown that for
water layers on some transition-metal surfaces the traditional
distinction between covalent and hydrogen bonds can be partially
or almost entirely lost~\cite{michaelides2010}. 
However, for the H$_2$O/MgO~(001) case
studied here, nuclear quantum effects turn out to have a rather small
net effect on the chemical potential, changing it by only 30~meV.
This is not because the the zero-point energies are small (they are not),
but because the intra-molecular vibrational frequencies of H$_2$O are
not greatly changed on adsorption, and the effects of the changes
are partly cancelled by the zero-point energies of the new molecule-surface
modes created from translational and rotational modes of the free molecule.
However, this will not be the case for many other water-adsorption systems,
and we believe there is ample scope for interesting applications
of our methods.

The practical calculations of this paper and paper~I were on the
H$_2$O/MgO~(001) system at very low coverage, where interactions
between adsorbed molecules can be ignored. However, we focused on this
low-density limit out of necessity rather than choice. In most
practical situations, intermolecular interactions play a major role, and certainly
have a strong effect on thermal desorption rates and on the surface
coverage in thermal equilibrium with a given gas-phase pressure.
For {\em ab initio} statistical mechanics, calculations away from
the low-density limit present two kinds of problem.
First, the simulated systems need to be larger, in order to reduce
system-size errors to an acceptable level. Second, memory times, and hence
sampling times usually become much longer, because molecules tend to
hinder each other's translational and rotational motion.
We know from our own work
with empirical interaction models for H$_2$O/MgO~(001)~\cite{fox2007,fox2009}
that, if we use the PMF approach, simulation times need to be 
at least 10 times as long at higher coverage than in the low-density limit. 
Nevertheless, with empirical interaction
models, the calculation of adsorption isotherms (coverage as a function
of chemical potential at fixed $T$) has been routinely practiced for 
many years, using grand canonical Monte Carlo 
(GCMC)~\cite{norman1969,frenkel2002} and other
techniques. It is clearly an important challenge for the immediate
future to develop the techniques that will allow the same to be
done with {\em ab initio} methods, and ultimately with {\em ab initio}
path-integral methods. This will not necessarily be done with GCMC, though,
because there may be more effective strategies.

Before concluding, we return briefly to the issue of improving {\em ab
initio} accuracy, which we noted in the Introduction. Improvements
are much needed, because DFT energetics of surface systems often
depends significantly on the approximation used for
exchange-correlation energy. Increasingly effective methods for
applying highly-correlated quantum chemistry to extended
systems~\cite{paulus2006,manby2006,li2008,nolan2009,marsman2009,harl2009},
the growing power of quantum Monte Carlo
techniques~\cite{alfe2006,gurtubay2007,drummond2007,pozzo2008,binnie2009,ma2009,sola2009},
and progress with improved van der Waals DFT
functionals~\cite{klimes10} give promising signs that {\em ab initio}
calculations on surface adsorbates with chemical accuracy or better
are now coming within reach. These improvements will enhance the
importance of being able to account for nuclear quantum corrections.

In conclusion, we have presented {\em ab initio} path-integral
techniques which allow the calculation of the chemical potential
of adsorbate molecules in thermal equilibrium, with the nuclei
treated fully quantum mechanically. We have verified that the
techniques give correct results by applying them to a set of models
where the quantum corrections are exactly known. We have demonstrated
the practical effectiveness of the techniques by applying them
to the H$_2$O molecule adsorbed on the MgO~(001) surface. We find
that in this case nuclear quantum corrections to the chemical  potential
are only $\sim 30$~meV, but we have noted the possibility of applying the
techniques to other systems where the corrections should be much larger.


\section*{Acknowledgements}
\label{sec:ack}

The work was supported by allocations of time on the HPCx
service provided by EPSRC through the Materials Chemistry Consortium
and the UK Car-Parrinello Consortium, and by resources provided
by UCL Research Computing. The work was conducted as part of
a EURYI scheme award as provided by EPSRC (see www.esf.org/euryi).



\section*{Appendix: Test by adsorption on the rigid surface}
\label{sec:app}

In Sec.~\ref{sec:testing}, we presented some tests of the path-integral
techniques applied to models in which carefully chosen external
potentials act on the molecule in free space. We summarise here a
different kind of test, in which full path-integral calculations
of the mean force as a function of height $z$ are used to calculate
the chemical potential for surface adsorption, but with the MgO slab
held rigid. These calculations are of exactly the same kind as the full
calculations reported in Sec.~\ref{sec:full_calcs}, in that the entire
system (MgO slab + H$_2$O molecule) is treated by DFT. However,
we modify the system to ensure that it
is closely harmonic with the molecule on the surface by adding
artificial potentials of the kind already used in Sec.~\ref{sec:testing};
this enables us to obtain quasi-exact values for the difference
between the quantum and classical values of the chemical potential,
which can be used to validate the path-integral methods.

The potentials that we have added to the DFT total energy to ensure
almost harmonic behaviour have the form
$V_{\rm ext} + W_{\rm ext} + K_{\rm ext}$. Here,
$V_{\rm ext}$ is the potential acting on the normal to the molecular
plane (Sec.~\ref{sec:normal_to_plane}), and $W_{\rm ext}$ is
the potential acting on the molecular bisector (Sec.~\ref{sec:other_tests}). 
The potential $K_{\rm ext}$
is a harmonic potential confining the motion of the molecular
centre of mass in the plane parallel to the surface:
\begin{equation}
K_{\rm ext} = \frac{1}{2} \beta ( ( x^{\rm cm} )^2 + 
( y^{\rm cm} )^2 ) \; ,
\end{equation}
with ${\bf r}^{\rm cm}$ the centre-of-mass position defined in
Sec.~\ref{sec:c_of_m}. The $V_{\rm ext}$ potential is zero when the normal
to the molecular plane points along the $z$-axis, which in the present
context means the normal to the MgO~(001) surface. The $W_{\rm ext}$
potential is zero when the molecular bisector points along the $x$-axis,
and we take this to be a cubic axis of the crystal parallel to the
surface. Finally, we take the $K_{\rm ext}$ potential to be zero
at a point near a surface Mg ion. All three
potentials $V_{\rm ext}$, $W_{\rm ext}$ and $K_{\rm ext}$ are invariant
under translation of the molecule along the $z$-direction, and they
act on the molecule in exactly the same way when it is far from the
surface as when it is on the surface. The values of the
spring constants $\alpha$, $\gamma$ and $\beta$ are chosen to
be 2.00~eV, 2.00~eV and 10.0~eV/\AA$^2$ respectively. From the harmonic
vibration frequencies of the molecule on the surface and in free
space, we easily derive the chemical potential
$\Delta \mu^\dagger$ at any temperature $T$, using either classical
or quantum theory. At $T = 100$~K, we find that the difference
$( \Delta \mu^\dagger )_{\rm qc}$ (quantum minus classical) 
has the small value 9~meV.

As in Sec.~\ref{sec:full_calcs}, the potential of mean force calculations
were performed with 8 beads, and the results were then corrected by
thermodynamic integration with respect to number of beads, going from
8 to 16 to 32 to 64 beads. For comparison, the PMF calculations
were also performed with classical nuclei. We find that with 8 beads
the quantum-classical difference $( \Delta \mu^\dagger )_{\rm qc}$ is
$- 13 \pm 1$~meV, which differs substantially from the quasi-exact
value of 9~meV. However the corrections from thermodynamic integration
over the number of beads are large, and they do not cancel between
gas and surface. In the gas phase, the change of free energy on going
from 8 to 64 heads is $198 \pm 3$~meV, while on the surface the
change is $180 \pm 4$~meV. This means that going from 8 to 64 beads
stabilises the molecule on the surface by 18~meV, and the
resulting corrected value for $( \Delta \mu^\dagger )_{\rm qc}$ comes
to $5 \pm 5$~meV, which agrees well with the quasi-exact value within
statistical errors.

\end{document}